\documentclass{article}
\usepackage{wds10,epsf}
\usepackage{multirow}
\usepackage{multicol}
\newcommand{\konst}{\mathop{\rm const.}\nolimits}
\newcommand{\dif}{\mathrm{d}}
\newcommand{\RS}{{\cal R}}

\newcommand{\DS}{\Delta}
\newcommand{\Lz}{\Lambda}

\lefthead{Hru\v{s}ka and Podolsk\'{y}}
\righthead{Kundt spacetimes in the Gauss--Bonnet Gravity}

\begin{document}

\title{Kundt Spacetimes in the Gauss--Bonnet Gravity}

\author{O. Hru\v{s}ka and J. Podolsk\'{y}}

\affil{Charles University in Prague, Institute of Theoretical Physics, Faculty  of  Mathematics  and  Physics,
     Prague, Czech Republic.}

\begin{abstract}
The Gauss--Bonnet gravity is a special case of so-called Quadratic Gravity, which is an extension of Einstein's theory with additional terms in action that are quadratic combinations of the Riemann tensor and its contractions. These corrections are needed, for example, in perturbative quantum gravity. We consider the family of Kundt spacetimes, which is defined in a purely geometrical way by admitting a shear-free, twist-free and expansion-free null geodesic congruence. In particular, we focus on the Kundt solutions without gyratonic terms, and we investigate the constraints  imposed by the Einstein--Gauss--Bonnet field equations. The conditions for the metrics to be of various algebraic types are also studied.
\end{abstract}

\begin{article}

\section{Introduction}
In Quadratic Gravity, the action contains additional terms to the Einstein--Hilbert action that are quadratic in curvature. It has the form (see for example [\markcite{{\it Deser and Tekin}, 2003}] or [\markcite{{\it M\'{a}lek and Pravda}, 2011}])
\begin{eqnarray}
\textstyle S=\int \dif^D x\sqrt{-g}\left(\frac{1}{\kappa}(R-2\Lz)+\alpha\,R^2+\beta\,R_{ab}^2+\gamma(R_{abcd}^2-4R_{ab}^2+R^2)\right),
\end{eqnarray}
where $D$ is the dimension of the spacetime. The Gauss--Bonnet gravity, see [\markcite{{\it Lovelock}, 1971}], [\markcite{{\it Zwiebach}, 1985}] is obtained by setting ${\alpha=0=\beta}$, in which case the field equations are of the second order. Indeed, variation of the action with respect to the metric yields the vacuum field equations
\begin{eqnarray}\label{eq:QG}\textstyle
&&\textstyle\frac{1}{\kappa}\left(R_{ab}-\frac{1}{2}R\,g_{ab}+\Lz\,g_{ab}\right)\\
&&\textstyle+2\gamma\left(R\,R_{ab}-2R_{abcd}\,R^{cd}+R_{acde}\,R_b^{\,\,cde}-2R_{ac}R_b^{\,\,c}-\frac{1}{4}g_{ab}(R_{cdef}^2-4R_{cd}^2+R^2)\right)=0.\nonumber
\end{eqnarray}
For ${\gamma=0}$, Einstein's equations are recovered. It can also be shown that in four dimensions (${D=4}$), the additional Gauss--Bonnet term does not contribute to the field equations.\\
$\bullet$\textit{The Kundt geometry} is defined by the property that it admits a non-expanding, non-twisting and shear-free null geodesic congruence. General Kundt metric in an arbitrary dimension (that is not restricted by field equations) is
\begin{eqnarray}\label{eq:Kundt}
\dif s^2=g_{ij}(u,x)\,\dif x^i \dif x^j+2g_{ui}(r,u,x)\,\dif u \dif x^i-2\dif u\,\dif r+g_{uu}(r,u,x)\,\dif u^2,
\end{eqnarray}
where $r$ is the affine parameter along the optically privileged vector field ${\mathbf{k}=\partial_r}$ that is everywhere tangent to the null surfaces ${u=\konst}$. The remaining coordinates ${x\equiv(x^2,x^3,\ldots,x^{D-1})}$ cover the transverse Riemannian space. The Riemann tensor, Ricci tensor and Ricci scalar for this transverse space with the metric $g_{ij}$ will be labeled as $\RS_{ijkl}$, $\RS_{ij}$ and $\RS$, respectively. It was shown in [\markcite{{\it Podolsk\'{y} and \v{S}varc}, 2015}], that for algebraically special Kundt spacetimes (for which $\mathbf{k}$ is at least double degenerate Weyl-aligned null direction (WAND)), the metric component $g_{ui}$ has the form ${g_{ui}=e_i(u,x)+f_i(u,x)r}$.
In this contribution, we restrict our attention on the simpler case (without gyratons)
\begin{eqnarray}\label{eq:gui0}
g_{ui}=0.
\end{eqnarray}
Explicit components of the Riemann and Ricci tensors $R_{abcd}$ and $R_{ab}$ for (\ref{eq:Kundt}) have already been calculated in [\markcite{{\it Podolsk\'{y} and \v{S}varc}, 2015}] and [\markcite{{\it Podolsk\'{y} and \v{S}varc}, 2016}], and we will employ them here.\\
$\bullet$\textit{The Einstein space} is defined by the property that the Ricci tensor is proportional to the metric: ${R_{ab}=\textstyle\frac{1}{D}R\,g_{ab}}$, we will examine one of the special cases with a transverse Einstein space, where
\begin{eqnarray}
\RS_{ab}=\textstyle\frac{1}{D-2}\RS\,g_{ab}.
\end{eqnarray}
$\bullet$\textit{In the special case of transverse space with constant spatial curvature}, the Ricci scalar $\RS$ does not depend on the spatial coordinates $x^i$, so it can only depend on $u$. The Riemann tensor, Ricci tensor and their squares are then (see for example [\markcite{{\it Griffiths and Podolsk\'{y}}, 2009}])
\begin{eqnarray}\label{eq:ccR}
\RS_{ijkl}=\textstyle\frac{\RS}{(D-3)(D-2)}\left(g_{ik}g_{jl}-g_{il}g_{jk}\right), \quad \RS_{ij}=\textstyle\frac{\RS}{D-2}g_{ij}, \quad
\RS_{ijkl}^2=\textstyle\frac{2\RS^2}{(D-3)(D-2)},\quad \RS_{ij}^2=\textstyle\frac{\RS^2}{D-2}.
\end{eqnarray}
The spatial metric can be written in the canonical form
\begin{eqnarray}\label{eq:gijcc}
\textstyle g_{ij}=P^{-2}\delta_{ij},\quad\textrm{where}\quad P=1+\frac{\RS}{4(D-3)(D-2)}\left((x^2)^2+\cdots+ (x^{D-1})^2\right).
\end{eqnarray}
\section{The field equations}
Now, we substitute the Kundt metric (\ref{eq:Kundt}) with (\ref{eq:gui0}) into the vacuum field equations (\ref{eq:QG}). We will explicitly evaluate and investigate each component, and derive the restrictions put on the Kundt metric.\\
Firstly, we will examine the \textit{general case} where we do not make any assumptions about the Riemann or Ricci tensors. Then we will focus on a special case of the \textit{Einstein space} with a specific Ricci scalar. Thirdly, we will focus on the case with \textit{constant spatial curvature}.\\
Both the $rr$ and $ri$ field components are trivial. The first non-trivial component is thus the $ru$-component:
\subsection{General case}
\subsubsection*{$ru$-component:}
The $ru$-component of the Gauss--Bonnet vacuum field equations is
\begin{eqnarray}\label{eq:ruGB}
\textstyle\frac{1}{2}\,\RS-\Lz+2\kappa\gamma\,\big(\frac{1}{4}\RS^2_{ijkl}-\RS^2_{ij}+\frac{1}{4}\RS^2\big)=0.
\end{eqnarray}
It puts no restriction on $g_{uu}$, but gives us constraint on $\RS^2_{ijkl}$, $\RS^2_{ij}$ and $\RS^2$ of the transverse space.
\subsubsection*{$ij$-component:}
Calculating the $ij$-component of the field equations and using (\ref{eq:ruGB}), we obtain
\begin{eqnarray}\label{eq:ijGB}
S_{ij}\,g_{uu,rr}+\RS_{ij}+2\kappa\gamma\left(\RS\,\RS_{ij}-2\RS_{ikjl}\,\RS^{kl}+\RS_{iklm}\,\RS_j^{\,\,\,klm}-2\RS_{ik}\,\RS_j^{\,\,\,k}\right)=0,
\end{eqnarray}
where
\begin{eqnarray}\label{eq:Sij}
S_{ij}\equiv\textstyle-\frac{1}{2}g_{ij}+2\kappa\gamma\,\big(\RS_{ij}-\frac{1}{2}\RS\,g_{ij}\big).
\end{eqnarray}
\textit{In this section}, we will \textit{assume} ${S_{ij}\neq 0}$. It is zero in a special case of the Einstein space that will be discussed in the next section. We can take a trace of (\ref{eq:ijGB}) and calculate $g_{uu}$, that can be written using (\ref{eq:ruGB}) as
\begin{eqnarray}\label{eq:guuGB}
g_{uu}=b\,r^2+c\,r+d,\quad \textrm{where}\quad b=\textstyle\frac{4\Lz-\RS}{D-2+2\kappa\gamma(D-4)\RS},
\end{eqnarray}
so $g_{uu}$ is at most quadratic in $r$. We see that ${\gamma=0}$ is equivalent to ${D=4}$, because the Gauss--Bonnet term does not contribute to the field equations in four dimensions, so in this case $g_{uu}$ is
\begin{eqnarray}
\textstyle g_{uu}=\frac{4\Lz-\RS}{D-2}r^2+c(u,x)r+d(u,x),
\end{eqnarray}
which agrees with previous results in Einstein's theory [\markcite{{\it Krtou\v{s} et al.}, 2012}]. If we substitute for $g_{uu,rr}$ of (\ref{eq:guuGB}) back into (\ref{eq:ijGB}), we obtain an equation for the spatial metric $g_{ij}$
\begin{eqnarray}\label{eq:ijGB1}
&&\Big(D-2+4\kappa\gamma(D-4)(1+\kappa\gamma\,\RS)\,\RS+16\kappa\gamma\Lz\Big)\RS_{ij}+(1+2\kappa\gamma\,\RS)(\RS-4\Lz)g_{ij}\\
&&\quad\,\,+2\kappa\gamma\big(D-2+2\kappa\gamma(D-4)\RS\big)\big(-2\RS_{ikjl}\,\RS^{kl}+\RS_{iklm}\,\RS_j^{\,\,\,klm}-2\RS_{ik}\,\RS_j^{\,\,\,k}\big)=0.\nonumber
\end{eqnarray}
\subsubsection*{$ui$-component:}
The $ui$-component of the Gauss--Bonnet field equations has the form
\begin{eqnarray}\label{eq:uiGB}
\textstyle S^{ij}(g_{uu,rj}-2g^{lm}g_{l[j,u||m]})+2\kappa\gamma\big(-2g^{ij}\RS^{kl}+\RS^{ikjl}\big)g_{k[j,u||l]}=0,
\end{eqnarray}
where ``$_{||}$'' denotes spatial covariant derivative, for example ${g_{uu||ij}=g_{uu,ij}-\Gamma^k_{ij}g_{uu,k}}$. The non-transverse components of the metric behave like scalars under this derivative. We substitute from (\ref{eq:guuGB}) to get the equation
\begin{eqnarray}\label{eq:uiGB1}
S^{ij}(2r\,b_{,j}+c_{,j}-2g^{lm}g_{l[j,u||m]})+2\kappa\gamma\big(-2g^{ij}\RS^{kl}+\RS^{ikjl}\big)g_{k[j,u||l]}=0.
\end{eqnarray}
Equation in the first power of $r$ gives us
\begin{eqnarray}\label{eq:uiGBr}
\textstyle S^{ij}\,b_{,j}=0,
\end{eqnarray}
which means, thanks to (\ref{eq:guuGB}),
\begin{eqnarray}
\textstyle \frac{D-2+8\kappa\gamma\Lz(D-4)}{(D-2+2\kappa\gamma(D-4)\RS)^2}S^{ij}\RS_{,j}=0.
\end{eqnarray}
We assume that the denominator is not zero here (it is zero in the special case mentioned in the next section). This means that either
\begin{eqnarray}\label{eq:uiGBr1}
\textstyle 8\kappa\gamma\Lz=-\frac{D-2}{D-4},\quad \textrm{or}\quad S^{ij}\RS_{,j}=0,
\end{eqnarray}
where the first option means ${b=\konst}$ from (\ref{eq:guuGB}) and it does not depend on $\RS$. The simplest solution for the second option would be ${\RS_{,i}=0}$, but that is not the most general solution.\\
Equation in the zeroth power of $r$ in (\ref{eq:uiGB1}) gives us the constraint
\begin{eqnarray}\label{eq:uiGBr0}
\textstyle S^{ij}(c_{,j}-2g^{lm}g_{l[j,u||m]})+2\kappa\gamma\big(-2\RS^{kl}g^{ij}+\RS^{ikjl}\big)g_{k[j,u||l]}=0.
\end{eqnarray}
\subsubsection*{$uu$-component:}
Finally, we calculate the $uu$-component of the field equations and substitute for the $ru$-equation, which yields
\begin{eqnarray}\label{eq:uuGB}
&&\textstyle S^{ij}\big(g_{uu||ij}+g_{ij,uu}-\frac{1}{2}g_{uu,r}g_{ij,u}-\frac{1}{2}g^{kl}g_{ki,u}g_{lj,u}\big)\nonumber\\
&&\quad\,\,\,+2\kappa\gamma\,g^{ij}g_{k[i,u||l]}g_{m[j,u||n]}\left(g^{km}g^{ln}-2g^{kl}g^{mn}\right)=0.
\end{eqnarray}
We will now use (\ref{eq:guuGB}) to write
\begin{eqnarray}\label{eq:uuGB1}
&&\textstyle S^{ij}\left[r^2\,b_{||ij}+r(c_{||ij}-b\,g_{ij,u})
+d_{||ij}-\frac{1}{2}c\,g_{ij,u}+g_{ij,uu}-\frac{1}{2}g^{kl}g_{ki,u}g_{lj,u}\right]\nonumber\\
&&\qquad\qquad\qquad\qquad\qquad\quad\!+2\kappa\gamma\, g^{ij}\left(g^{km}g^{ln}-2g^{kl}g^{mn}\right)g_{k[i,u||l]}g_{m[j,u||n]}=0.
\end{eqnarray}
Equation in the second power of $r$ is
\begin{eqnarray}\label{eq:uuGBr2}
\textstyle S^{ij}\,b_{||ij}=0.
\end{eqnarray}
If we make a covariant derivative of (\ref{eq:uiGBr}) and substitute for (\ref{eq:uuGBr2}), we obtain
\begin{eqnarray}
\left(g^{ij}\,\RS_{,i}-2\RS^{ij}_{\,\,\,\,||i}\right)b_{,j}=0,
\end{eqnarray}
where we can substitute for $b$ from (\ref{eq:guuGB}), and, provided that $b$ is not a constant, we obtain
\begin{eqnarray}\label{eq:uuGBr2a}
g^{ij}\,\RS_{,i}\,\RS_{,j}=2\RS^{ij}_{\,\,\,\,||i}\,\RS_{,j}.
\end{eqnarray}
Equation in the first power of $r$ in (\ref{eq:uuGB1}) is
\begin{eqnarray}\label{eq:uuGBr1}
\textstyle S^{ij}\left(c_{||ij}-b\,g_{ij,u}\right)=0.
\end{eqnarray}
Again, the simplest, but not the most general solution, is ${c_{||ij}=b\,g_{ij,u}}$.\\
Finally, the equation in the zeroth power of $r$ in (\ref{eq:uuGB1}) is
\begin{eqnarray}\label{eq:uuGBr0}
\textstyle S^{ij}\big(d_{||ij}+g_{ij,uu}-\frac{1}{2}c\,g_{ij,u}-\frac{1}{2}g^{kl}g_{ki,u}g_{lj,u}\big)+2\kappa\gamma\, g^{ij}\left(g^{km}g^{ln}-2g^{kl}g^{mn}\right)g_{k[i,u||l]}g_{m[j,u||n]}=0.
\end{eqnarray}

\subsection{Special Einstein space (${S_{ij}=0}$)}
We will now consider the special case of the Einstein space, where ${S_{ij}=0}$, of (\ref{eq:Sij}). This means that
\begin{eqnarray}\label{eq:RSijGB}
\textstyle\RS_{ij}=-\frac{1}{2\kappa\gamma(D-4)}\,g_{ij}=\frac{1}{D-2}\RS \,g_{ij},\quad \RS=-\frac{D-2}{2\kappa\gamma(D-4)}, \quad \RS_{ij}^2=\frac{\RS^2}{D-2}.
\end{eqnarray}
Let us note that this is \textit{not the general case of an Einstein space}, but it is interesting due to the fact that neither of the field equations will give us a restriction on the $g_{uu}$ component of the metric, as we will see below.
\subsubsection*{$ru$-component:}
We can substitute for $\RS_{ij}$ and $\RS$ from (\ref{eq:RSijGB}) into (\ref{eq:ruGB}) and we will get a constraint
\begin{eqnarray}\label{eq:ruGBE}
\RS_{ijkl}^2=\textstyle\left(\frac{D-2}{2\kappa\gamma(D-4)}\right)^2+\frac{4\Lambda}{2\kappa\gamma}.
\end{eqnarray}
\subsubsection*{$ij$-component:}
If we substitute (\ref{eq:RSijGB}) into (\ref{eq:ijGB}), we will receive an equation
\begin{eqnarray}\label{eq:ijGBE}
\textstyle\RS_{iklm}\,\RS_j^{\,\,klm}=\frac{2}{(2\kappa\gamma(D-4))^2}\,g_{ij}.
\end{eqnarray}
Now, we can make a trace of this equation and substitute for (\ref{eq:ruGBE}) to obtain
\begin{eqnarray}\label{eq:ijGBEtr}
\textstyle8\kappa\gamma\Lz=-\frac{D-2}{D-4}.
\end{eqnarray}
This equation links together the constants of the theory and the dimension. Since the Gauss--Bonnet gravity is relevant only for ${D>4}$, we see that $\Lz$ has the opposite sign as $\kappa\gamma$. Notice that equation (\ref{eq:ijGBEtr}) coincides with the first option in equation (\ref{eq:uiGBr1}).
\subsubsection*{$ui$-component:}
Again, the term $S_{ij}$ in (\ref{eq:uiGB}) is zero and we have the equation for the spatial metric
\begin{eqnarray}\label{eq:uiGBE}
\textstyle \left(\frac{1}{\kappa\gamma(D-4)}g^{kl}g^{ij}+\RS^{ikjl}\right)g_{k[j,u||l]}=0.
\end{eqnarray}
It can be seen that the term in the brackets cannot be zero, because a contraction of that term would give us $\RS_{ij}$ inconsistent with (\ref{eq:RSijGB}).
\subsubsection*{$uu$-component:}
For ${S_{ij}=0}$ the $uu$-equation reduces to
\begin{eqnarray}\label{eq:uuGBE}
g^{ij}\left(g^{km}g^{ln}-2g^{kl}g^{mn}\right)g_{k[i,u||l]}g_{m[j,u||n]}=0,
\end{eqnarray}
so we have a specific condition for the spatial metric. We again see that in the case of the Einstein space, neither of the field equations gives us a restriction on $g_{uu}$.

\subsection{Constant spatial curvature}
Here, we focus on the special situation when the spatial metric has constant curvature, see (\ref{eq:ccR}). While this is a special case of a general Einstein space, it is \textit{not a subcase of the Einstein space that we investigated in the previous section} because, as we will see below, the Ricci tensor will be different from (\ref{eq:RSijGB}).
\subsubsection*{$ru$-component:}
When the spatial metric $g_{ij}$ has a constant curvature, we can substitute for $\RS_{ijkl}$, $\RS_{ij}$ and $\RS$ from (\ref{eq:ccR}) into (\ref{eq:ruGB}) and we get the equation
\begin{eqnarray}\label{eq:ruGBcc}
\textstyle\kappa\gamma\frac{(D-4)(D-5)}{(D-3)(D-2)}\,\RS^2+\RS-2\Lz=0,
\end{eqnarray}
which is a quadratic equation for $\RS$. Since the coefficients are constant, $\RS$ must also be a constant, therefore independent of $u$, and both solutions are different from (\ref{eq:RSijGB}). This also means that the metric components $g_{ij}$ are also constants, see (\ref{eq:gijcc}).
\subsubsection*{$ij$-component:}
Since $\RS$ in (\ref{eq:ruGBcc}) is different from $\RS$ in (\ref{eq:RSijGB}), the term $S_{ij}$ in (\ref{eq:ijGB}) will remain non-zero and the metric component $g_{uu}$ has the same form as (\ref{eq:guuGB}) with the exception that $b$ is now constant, because $\RS$ is constant.
\subsubsection*{$ui$-component:}
Since the metric component $g_{ij}$ does not depend on $u$, the equation (\ref{eq:uiGBr0}) reduces to
\begin{eqnarray}\label{eq:uiGBcc}
c_{,i}=0,
\end{eqnarray}
so $c$ can at most depend on $u$.
\subsubsection*{$uu$-component:}
Because ${g_{ij,u}=0}$ here, the $uu$-component (\ref{eq:uuGB}) greatly simplifies and if we substitute from the previous equations, we get an equation
\begin{eqnarray}\label{eq:uuGBcc}
g^{ij}d_{||ij}\equiv\DS d=0.
\end{eqnarray}
Since $d$ can be any function of $u$, this may represent a profile of a gravitational wave. While the field equations determine $b$ and give restriction on the spatial dependence of $c$ and $d$, the dependence of these functions on $u$ remains arbitrary, so that any profile of the wave can be prescribed.
\section{Determining algebraical types}
We can now insert our results into the equations that determine the conditions for respective algebraical types of the Weyl tensor derived in [\markcite{{\it Podolsk\'{y} and \v{S}varc}, 2015}]. The algebraic classification is based on [\markcite{{\it Coley et al.}, 2004}]. We will discuss all the three cases together, because as we will see, the general case and the case ${S_{ij}=0}$ will lead to similar equations. For the constant spatial curvature space, if the condition on type II(a) is fulfilled, then the remaining conditions are automatically fulfilled from the field equations.
\subsubsection{Type II(a)}
The condition for the algebraical subtype II(a) is
\begin{eqnarray}\label{eq:IIa}
\textstyle g_{uu,rr}=-\frac{2}{(D-2)(D-3)}\,\RS.
\end{eqnarray}
$\bullet$ For ${S_{ij}\neq0}$, we substitute for $g_{uu}$  from (\ref{eq:guuGB}) and we receive an equation
\begin{eqnarray}\label{eq:PGBgen}
2\kappa\gamma(D-4)\,\RS^2-(D-2)(D-4)\,\RS+(D-2)(D-3)4\Lambda=0,
\end{eqnarray}
which is a quadratic equation for $\RS$, which is constant.\\
$\bullet$ For ${S_{ij}=0}$, we substitute for $\RS$ from (\ref{eq:RSijGB}) and we get a condition for the metric component $g_{uu}$ (that is not limited by the field equations), namely
\begin{eqnarray}\label{eq:guuGBE}
g_{uu}=\tilde{b}\,r^2+\tilde{c}\,r+\tilde{d},\quad \textrm{where} \quad \tilde{b}=\textstyle\frac{1}{2\kappa\gamma(D-3)(D-4)},
\end{eqnarray}
where the functions $\tilde{b},\,\tilde{c},\,\tilde{d}$ are \textit{different} from the case (\ref{eq:guuGB}).\\
Thus we see that in the algebraical subtype II(a), the metric is \textit{at most quadratic} in $r$ in both cases, but with different coefficients and different scalar curvature. Hence, we can discuss the cases ${S_{ij}=0}$ and ${S_{ij}\neq0}$ together, without substituting for $\RS$ and $b,\,c,\,d$.\\
$\bullet$ For the case of a constant spatial curvature, the quadratic equation (\ref{eq:ruGBcc}) has different coefficients than (\ref{eq:PGBgen}), so comparing these two equations gives us a constraint on the constants $\kappa,\,\gamma,\,\Lambda$ and dimension $D$ (we assume ${D\neq 4}$):
\begin{eqnarray}
\textstyle\kappa\gamma\Lambda\left(8\kappa\gamma\Lambda+\frac{(D-2)^2(D^2-7D+14)}{(D-3)(D-4)^3}\right)=0,
\end{eqnarray}
so either
\begin{eqnarray}\label{eq:ccIIa}
\gamma\Lambda=0,\quad \textrm{or} \quad 8\kappa\gamma\Lambda=\textstyle-\frac{(D-2)^2(D^2-7D+14)}{(D-3)(D-4)^3}.
\end{eqnarray}
\subsubsection{Type II(b)}
The algebraical subtype II(b) is equivalent to the condition
\begin{eqnarray}\label{eq:IIb}
\RS_{ij}=\textstyle\frac{1}{D-2}\,\RS\,g_{ij},
\end{eqnarray}
i. e., the spatial part of the metric is the Einstein space, where $\RS$ is given either by (\ref{eq:RSijGB}), or (\ref{eq:PGBgen}), since (\ref{eq:IIb}) gives no restriction on $\RS$.
\subsubsection{Type II(c)}
The algebraical type II(c) gives us the condition
\begin{eqnarray}\label{eq:IIc}
{\cal C}_{ijkl}=0,
\end{eqnarray}
where ${\cal C}_{ijkl}$ represents the Weyl tensor of the spatial metric $g_{ij}$.
\subsubsection{Type II(d)}
Since $g_{ui}$ is zero, the metric is always of subtype II(d), because the corresponding condition of [\markcite{{\it Podolsk\'{y} and \v{S}varc}, 2015}] is automatically fulfilled.
\subsubsection{Type III(a)}
The algebraical type III is equivalent to II(abcd). In order to get the subtype III(a), we must have
\begin{eqnarray}\label{eq:IIIa}
\textstyle c_{,i}=\frac{-2}{D-3}g^{kl}g_{k[i,u||l]}.
\end{eqnarray}
\subsubsection{Type III(b)}
The algebraical subtype III(b) is equivalent to the condition
\begin{eqnarray}\label{eq:IIIb}
\textstyle g_{i[k,u||j]}-\frac{2}{D-3}(g_{ik}g^{lm}g_{l[m,u||j]}-g_{ij}g^{lm}g_{l[m,u||k]})=0.
\end{eqnarray}
\subsubsection{Type N}
In order to get the algebraical type N, all previous conditions II(abcd) and III(ab) must be fulfilled.
\subsubsection{Type O}
Finally, the condition for the conformally flat type O is
\begin{eqnarray}
&&\textstyle\left(r^2\,b_{||ij}+r(c_{||ij}-b\,g_{ij,u})+d_{||ij}+g_{ij,uu}-\frac{1}{2}c\,g_{ij,u}-\frac{1}{2}g^{kl}g_{ki,u}g_{lj,u}\right)\\
&&\textstyle-\frac{1}{D-2}g_{ij}\,g^{kl}\left(r^2\,b_{||kl}+r(c_{||kl}-b\,g_{kl,u})+d_{||kl}+g_{kl,uu}-\frac{1}{2}c\,g_{kl,u}-\frac{1}{2}g^{mn}g_{mk,u}g_{nl,u}\right)=0.\nonumber
\end{eqnarray}
Because ${b_{,i}=0}$, also ${b_{||ij}=0}$ so the equation in the second order of $r$ is automatically fulfilled.\\
Equation in the first order of $r$ is
\begin{eqnarray}\label{eq:Or1}
c_{||ij}-b\,g_{ij,u}=\textstyle\frac{1}{D-2}g_{ij}g^{kl}(c_{||kl}-b\,g_{kl,u}).
\end{eqnarray}
We can substitute for $b$ and $c$ and get a constraint for the spatial metric.\\
Equation in the zeroth order of $r$ is
\begin{eqnarray}\label{eq:Or0}
\textstyle d_{||ij}+g_{ij,uu}-\frac{1}{2}c\,g_{ij,u}-\frac{1}{2}g^{kl}g_{ki,u}g_{lj,u}=\textstyle\frac{1}{D-2}g_{ij}g^{kl}\Big(d_{||kl}+g_{kl,uu}-\frac{1}{2}c\,g_{kl,u}
-\frac{1}{2}g^{mn}g_{mk,u}g_{nl,u}\Big).
\end{eqnarray}
\section{Summary}
We have calculated explicitly the field equations (\ref{eq:QG}) for three cases: the general case ${S_{ij}\neq 0}$ (see (\ref{eq:Sij})), the case ${S_{ij}=0}$, and the case of a constant spatial curvature (\ref{eq:ccR}), (\ref{eq:gijcc}). The results are summarized in table 1.\\
\begin{table}[h]
\begin{center}
\begin{tabular}{|l|l|l|l|l|}
\hline
Case & $ru$ & $ij$ & $ui$ & $uu$\\
\hline
${S_{ij}\neq 0}$ & (\ref{eq:ruGB}) & (\ref{eq:guuGB}), (\ref{eq:ijGB1}) & (\ref{eq:uiGBr1}), (\ref{eq:uiGBr0}) & (\ref{eq:uuGBr2a}), (\ref{eq:uuGBr1}) (\ref{eq:uuGBr0})\\
\hline
${S_{ij}= 0}$ & (\ref{eq:ruGBE}) & (\ref{eq:ijGBE}), (\ref{eq:ijGBEtr}) & (\ref{eq:uiGBE}) & (\ref{eq:uuGBE})\\
\hline
CC & (\ref{eq:ruGBcc}) & (\ref{eq:guuGB}) with ${b=\konst}$ & (\ref{eq:uiGBcc}) & (\ref{eq:uuGBcc})\\
\hline
\end{tabular}
\caption{References to the results of respective components of the field equations for the three cases: ${S_{ij}\neq 0}$, ${S_{ij}= 0}$, and constant spatial curvature. The abbreviation ``CC'' stands for ``constant spatial curvature'' transverse space.}
\end{center}
\end{table}
\\
\\
We have also evaluated conditions for these three cases to be of various algebraic types. References to the conditions are given in table 2.\\
\begin{table}[h]
\begin{center}
\begin{tabular}{|l|l|l|l|l|l|l|l|l|}
\hline
Case & II(a) & II(b) & II(c) & II(d) & III(a) & III(b) & N & O\\
\hline
${S_{ij}\neq 0}$ & (\ref{eq:PGBgen}) & \multirow{2}*{(\ref{eq:IIb})} & \multirow{2}*{(\ref{eq:IIc})} & \multirow{2}*{always} & \multirow{2}*{(\ref{eq:IIIa})} & \multirow{2}*{(\ref{eq:IIIb})} & \multirow{2}*{III(a) and III(b)} & \multirow{2}*{(\ref{eq:Or1}) and (\ref{eq:Or0})}\\
${S_{ij}= 0}$ & (\ref{eq:guuGBE}) &  &  &  &  &  &  & \\
\hline
CC & \multicolumn{8}{|c|}{(\ref{eq:ccIIa})}\\
\hline
\end{tabular}
\caption{References to the conditions for the three cases: ${S_{ij}\neq 0}$, ${S_{ij}= 0}$, and constant spatial curvature to be of various algebraical types.}
\end{center}
\end{table}

\section{Conclusion}
We investigated the Kundt spacetimes without gyratons (those admitting expansion-free, shear-free and twist-free null geodesic congruence) in the Gauss--Bonnet gravity. We calculated all components of the field equations and wrote the equations that give restrictions on the algebraically special Kundt metric. We studied the general case, where the field equations dictate that the metric is at most quadratic in the affine coordinate $r$. Then we investigated the special case of Einstein spaces with Ricci scalar given by (\ref{eq:RSijGB}), where the field equations give no restriction on the metric component $g_{uu}$, and we also investigated the special case with constant spatial curvature, where the field equations are considerably simplified: we received wave equation for the part of $g_{uu}$ that does not depend on $r$. Such exact solutions represent gravitational waves. We also investigated the conditions for the metric for specific algebraical types, that are based on key Weyl scalars. Namely, we found out that if the constant curvature transverse space is of type II(a), it is automatically conformally flat.

\acknowledgments 
{The  work  was  supported by the Charles University Grant GAUK number 196516, SVV-260320 and GA\v{C}R P203/12/0118.}

\end{article}

\end{document}